\documentclass[prl,twocolumn,superscriptaddress]{revtex4-1}
\usepackage{amsfonts}
\usepackage{mathrsfs}
\usepackage{amsmath}
\usepackage{amssymb}
\usepackage[medium]{titlesec}
\usepackage{bm}
\usepackage[normalem]{ulem}
\usepackage{extarrows}
\usepackage{slashed}
\usepackage{isodateo}
\usepackage{graphicx} 
\usepackage{xcolor}
\usepackage[utf8]{inputenc}
\usepackage[bookmarksnumbered=true,bookmarksopen=true]{hyperref}
 \hypersetup{colorlinks,%
             linkcolor=[rgb]{0,0.3,0.6}, %
             citecolor=[rgb]{0,0.3,0.6}, %
             urlcolor=[rgb]{0,0.3,0.6}}
\usepackage{indentfirst}
\renewcommand{\FR}[2]{\displaystyle\frac{\,{#1}\,}{#2}}
\newcommand{\fr}[2]{\mbox{$\frac{\,{#1}\,}{#2}$}}
\newcommand{\n}{\nonumber}

\graphicspath{{fig/}}

\usepackage[eulergreek]{sansmath}

\def\bge{\begin{equation}}
\def\ede{\end{equation}}
\def\bga{\begin{aligned}}
\def\eda{\end{aligned}}
\def\bgp{\begin{pmatrix}}
\def\edp{\end{pmatrix}}
\def\bgs{\begin{subequations}}
\def\eds{\end{subequations}}
\newcommand{\order}[1]{\mathcal{O}({#1})}
\def\di{{\mathrm{d}}}

\def\mb{\mathbf}

\def\pd{\partial}
\def\ld{{\mathscr{L}}}

\def\la{\langle}\def\ra{\rangle}

\def\to{\rightarrow}

\def\ii{\mathrm{i}}

\def\al{\alpha}

\def\de{\delta}
\def\ep{\epsilon}

\def\lam{\lambda}

\def\si{\sigma}

\def\Mp{M_{\text{Pl}}}

\begin{document} 

\title{Large-Field Inflation and the Cosmological Collider}

\author{Matthew Reece}
\email{mreece@g.harvard.edu}
\affiliation{Department of Physics, Harvard University, Cambridge, MA 02138, USA} 

\author{Lian-Tao Wang}
\email{liantaow@uchicago.edu}
\affiliation{Department of Physics, University of Chicago, Chicago, IL 60637, USA}

\author{Zhong-Zhi Xianyu}
\email{zxianyu@tsinghua.edu.cn}
\affiliation{Department of Physics, Tsinghua University, Beijing 100084, China}  
 
\begin{abstract}

Large-field inflation is a major class of inflation models featuring a near- or super-Planckian excursion of the inflaton field. We point out that the large excursion generically introduces significant scale dependence to spectator fields through inflaton couplings, which in turn induces characteristic distortions to the oscillatory shape dependence in the primordial bispectrum mediated by a spectator field. This so-called cosmological collider signal can thus be a useful indicator of large field excursions. We show an explicit example with signals from the ``tower states'' motivated by the swampland distance conjecture.  
\end{abstract}

\maketitle

\noindent{\bf Introduction.} Despite the success of cosmic inflation as a widely accepted scenario for the primordial universe, its microscopic model largely remains unclear at the moment, due to the large degeneracy when confronting models with data. While it is virtually impossible to pin down \emph{the} model of inflation, it is nevertheless possible to probe some general properties of inflation models by cosmological observations \cite{Planck:2018jri}. 

A notable property is the excursion range $\Delta\phi$ of the inflaton field from the moment when CMB-scale modes left the horizon, to the time when inflation stopped. Very roughly, models with $\Delta\phi\gtrsim \Mp\simeq 2.4\times 10^{18}$GeV are called large-field inflation, while models with $\Delta\phi\ll \Mp$ are called small-field inflation. It turns out that large and small field models have drastically different properties and consequences, making $\Delta\phi$ a useful classifier of inflation models. 

One particular feature of large-field inflation is its noticeable scale dependence, as shown by the well-known Lyth bound $\Delta\phi\sim\sqrt{2\ep}N_e\Mp$ in single field slow roll (SFSR) models \cite{Lyth:1996im}. Here $\ep$ is the first slow-roll parameter and $40\lesssim N_e\lesssim60$ is the $e$-folding number. A large $\Delta\phi$ $(\gtrsim \Mp)$ implies a not-so-small $\ep$, and in turn implies a relatively high inflation scale and a potentially observable tensor mode, since the tensor-to-scalar ratio $r=16\ep$ in SFSR models. This conclusion survives in a much wider range of models beyond SFSR~\cite{Baumann:2011ws, Mirbabayi:2014jqa}.

Recently there has been revived interest in large-field inflation, and in particular in whether a scalar field can consistently traverse a super-Planckian excursion within quantum gravity. In particular, the swampland distance conjecture (SDC) holds that a super-Planckian distance $\Delta \phi \gg \Mp$ always corresponds to a breakdown of low-energy effective field theory, with $\phi$ acting as a modulus controlling the mass of an infinite tower of light states (henceforth ``tower states'') with masses behaving as $m_\text{tower} \sim \exp(- O(1) \times \Delta \phi/\Mp)$ asymptotically~\cite{Ooguri:2006in, Klaewer:2016kiy}. This characteristic exponential behavior can be understood as a feedback effect from the loops of the tower states modifying the kinetic term of $\phi$~\cite{Heidenreich:2018kpg, Grimm:2018ohb}. A controlled model of inflation should exist well away from this asymptotic region, which poses a familiar challenge for constructing realistic examples in string theory~\cite{Dine:1985he}.

While large super-Planckian field excursions may be in tension with consistent quantum gravity, modest excursions $\Delta \phi \sim O(1) \times \Mp$ are less constrained~\cite{Scalisi:2018eaz}, and arise in concrete examples such as axion monodromy~\cite{Silverstein:2008sg, McAllister:2008hb, McAllister:2014mpa}. Although most of the tower states can remain heavy in such a scenario, it is still expected that there are some massive states whose masses can change significantly as $\phi$ evolves over a Planckian range (which can even improve the behavior of the potential~\cite{Dong:2010in}). Thus, mild and visible scale dependence is a very generic feature of large-field inflation. It may well trigger the onset of dramatic events such as the proliferation of light states, particle production if a mass evolves non-adiabatically~\cite{Flauger:2016idt}, and even phase transitions~\cite{An:2020fff,An:2022cce}. It is therefore desirable to have a more clear and direct probe of such scale dependence, as a generic indicator of large-field inflation.

In this Letter, we exploit the recently developed cosmological collider observables to probe  large-field inflation. The aim of cosmological collider physics is to look for heavy states with mass around or even much heavier than the inflation Hubble parameter $H$~\cite{Chen:2009zp, Noumi:2012vr, Arkani-Hamed:2015bza,Lee:2016vti,Chen:2016uwp,Chen:2016hrz}. Such states were produced in a fast evolving background during inflation and then source a characteristic oscillatory signal in a soft limit of the 3-point correlator of the curvature perturbation $\zeta$~\cite{Arkani-Hamed:2015bza}. One normally defines a scaleless and dimensionless shape function $\mathcal{S}(k_1,k_2,k_3)$ to describe the 3-point correlator, with $k_i~(i=1,2,3)$ the magnitude of the three external momenta. Then, the cosmological collider signal appears when $k_1\simeq k_2\gg k_3$, the so-called squeezed limit, where we can write, ignoring a factor that might depend on ${\mb k_1}\cdot {\mb k_3}$,
\begin{align}
\label{eq_signalshape}
  \mathcal{S}_\text{signal}(k_1,k_2,k_3) \sim A\Big(\FR{k_1}{k_3}\Big)^\al\sin\Big[\omega\log\Big(\FR{k_1}{k_3}\Big)+\de\Big].
\end{align}
 Here $A$ is the overall amplitude of the signal and $(\al,\omega,\de)$ describe the shape dependence, which are in principle measurable from future CMB/LSS/21cm observations \cite{Meerburg:2019qqi,Meerburg:2016zdz,MoradinezhadDizgah:2018ssw,Kogai:2020vzz} and also calculable from a given model \cite{Wang:2021qez,Tong:2021wai}. For instance, with a tree level exchange of a scalar field of mass $m>3H/2$, we have $\al=-1/2$ and $\omega=\sqrt{m^2/H^2-9/4}$. 
 
The main point we will make in this letter is that the scale dependence in large-field inflation can be particularly significant for a massive spectator state (such as tower states) and this will make observables such as $\al$ and $\omega$ further dependent on $k_1/k_3$. It is not new that slow-roll correction can introduce scale dependence of $\order{\ep,\eta}$ for inflation correlators. The new observation is that the direct inflaton coupling, such as the $\phi$-dependent mass of the tower states, can introduce scale dependence of $\order{\ep^{1/2}}$ and this can be a significant effect for large-field inflation. Therefore, measuring the $k_1/k_3$-dependence in $\al$ and $\omega$ can be a useful indicator of large-field inflation. 

The recently updated constraint $r\lesssim 0.03$ by BICEP/Keck \cite{BICEP:2021xfz} translates to a bound in SFSR models, $16 \ep\lesssim 0.03 $. But we stress that this constraint on $\ep$ does not directly apply when there are spectator fields. In particular, the spectator fields appearing in our cosmological collider signals can suppress $r$ by introducing more power to the scalar mode than to the tensor \cite{An:2017hlx,Iyer:2017qzw}. We will show below that the spectator field with time-dependent mass could bring a new correction to $n_s$ in either direction, in addition to the $n_s$ correction from constant-mass spectator fields considered in \cite{An:2017hlx}.  Thus, in principle, the observational constraints on $n_s$ and $r$ need to be reanalyzed for our model. At the same time, Ref.~\cite{An:2017hlx} shows that the relative correction to $r$ is typically of $\order{1}$ in the parameter space we are interested in.  Since our aim is to illustrate a physical effect instead of fitting a specific model to data, we will loosely take $16\ep<\order{0.1}$.

\noindent{\bf Scale dependent inflation parameters.} The inflationary spacetime is close to but not exactly dS. The Hubble parameter $H=H(t)$ has a weak time dependence characterized by the slow-roll parameters, including: 
\begin{align}
  &\ep=-\FR{\dot H}{H^2},  &&\eta=\FR{\dot\ep}{H\ep}.
\end{align}
In single-field inflation models with an inflaton $\phi$, the inflaton is rolling with a nearly but not exactly constant speed $\dot\phi_0=\dot\phi_0(t)$, and the time dependence here is again characterized by the slow-roll parameters. The time dependence in $\phi_0(t)$, $\dot\phi_0(t)$, $H(t)$ can easily introduce a time-dependent mass to a spectator field $\si$. So time-dependent masses in inflation are rather generic. 
 
Therefore, choosing a reference time $t_*=0$,
\begin{align}
  H(t)=H_*\big[1-\ep_* H_* t+(\ep_*^2-\fr{1}{2}\ep_*\eta_*)(H_*t)^2+\cdots\big].
\end{align}
Using the relation $\dot\phi_0^2\simeq 2\ep \Mp^2H^2$, we get
\begin{align}
  \dot\phi_0(t)=\dot\phi_{0*}\big[1+(\fr{1}{2}\eta_*-\ep_* )H_*t+\cdots\big].
\end{align}
Using these relations we can find the time dependence   of the masses induced by some familiar interactions. For example, the spectator can have a nonminimal coupling to the Ricci scalar $-\xi R\si^2$. Then using $R=12H^2+6\dot H$ we have $m_\si^2=m_{\si*}^2(1-2\ep_* H_* t+\cdots)$ with $m_{\si*}^2=24\xi H_*^2$.  Take another example of $(\pd\phi)^2\si^2/\Lambda^2$; we have $m_\si^2=m_{\si*}^2[1+(\eta_*-2\ep_*)H_*t+\cdots]$ with $m_{\si*}^2=2\dot\phi_{0*}^2/\Lambda^2$.
 
In all such examples the time dependence is weak, in the sense of being $\order{\ep,\eta}$. Therefore, to see an $\order{1}$ effect from this time dependence, we would need the number of e-folds $N\simeq \min\{1/\ep,1/\eta\}\sim \order{30}$. Practically we have access only to $\order{10}$ $e$-folds from CMB and LSS. So it is difficult to see this time dependence directly.

Stronger time dependence is possible if we consider non-derivative couplings of the form $m_{\si*}^2 f(\phi/\Mp) \si^2$. Such a coupling breaks the approximate shift symmetry of the inflaton field, which is expected to be broken at least by quantum gravity effects. The evolution of the inflaton field leads to $m^2_\si(t) = m_{\si*}^2 f'  (1-\sqrt{2\ep_*} H_* t+\cdots)$.  In this case  we only need $\order{\ep^{-1/2}}$ $e$-folds to see the effect. For large-field inflation the required $e$-folds could be $\order{10}$, and thus possible for CMB/LSS observation. We note that this coupling is small, which would not change the dynamics of the inflaton. For the same reason, it would not generate a sizable non-Gaussianity signal. At the same time, there could certainly be other, shift-symmetry preserving, couplings which can generate an observable signal. The effect of this non-derivative coupling would be modulating the shape of the signal, as we will discuss in detail below.

For large-field models, a rather generic interaction exists, namely the ``tower states'' motivated by the SDC, with mass $\propto e^{-\al\phi/\Mp}$, where $\al$ is an $\order{1}$ number. With rolling background $\phi(t)$, this introduces a time-dependent mass $m_\si^2=m_{\si*}^2e^{-2\al\sqrt{2\ep_*}Ht}$ which is again an $\order{\ep^{1/2}}$ effect. Although this exponential ansatz is best motivated in asymptotic regions of the potential, the scale-dependent frequency and size of the resulting signal is largely independent of the particular ansatz we choose, so long as the spectator couples directly to the inflaton via non-derivative couplings. Hence, we focus on this particular ansatz as an example, and present results for an alternative choice of direct spectator-inflaton coupling in the appendix.

\noindent{\bf Quasi-single (large) field inflation.} 
As a concrete example we study quasi-single field inflation. In its original form \cite{Chen:2009zp}, we have a massive scalar field $\si$ with mass $m_\si\sim\order{H}$ coupled to the inflaton through a derivative coupling $(\pd_\mu\phi)^2\si$. At the same time,  $\si$ has a cubic self-coupling $\lam\si^3$ as well. The point is that the $\si$ self-coupling is not constrained by the slow-roll condition and can a priori be large (namely $\lam/H\sim 1$ or larger). So the bispectrum from this cubic coupling can be potentially large. This is a well studied model with large non-Gaussianity. Below, we will consider the effect of inflaton modulation in this model. 

As discussed before, we assume that the inflaton modulates the mass $m_\si$ through an exponential factor. Then, the Lagrangian is given by
\begin{align}
\label{eq_ActionTD}
  \ld=& -\fr{1}{2}(\pd_\mu\phi)^2-V(\phi)-\fr{1}{2}(\pd_\mu\si)^2-\fr{1}{2}e^{-2\al\phi/\Mp}m^2\si^2\n\\
  &-\fr{1}{6}\lam_3\si^3+\fr{1}{2}\lam_5 (\pd_\mu\phi)^2\si.
\end{align}
The model described by the Lagrangian (\ref{eq_ActionTD}) is certainly not complete, as the $\si$ potential is not stable, and the inflaton rolling introduces a tree-level tadpole to $\si$. But we can still treat it as an effective description of a more complete model, in which the $\si$ field is stabilized at some background value $\si_0$, for example, by an approximate cancellation between the terms proportional to  $\lambda_3$ and $\lambda_5$, and we are simply expanding the full action around $\si_0$.

\begin{figure}
  \centering
  \includegraphics[width=0.42\textwidth]{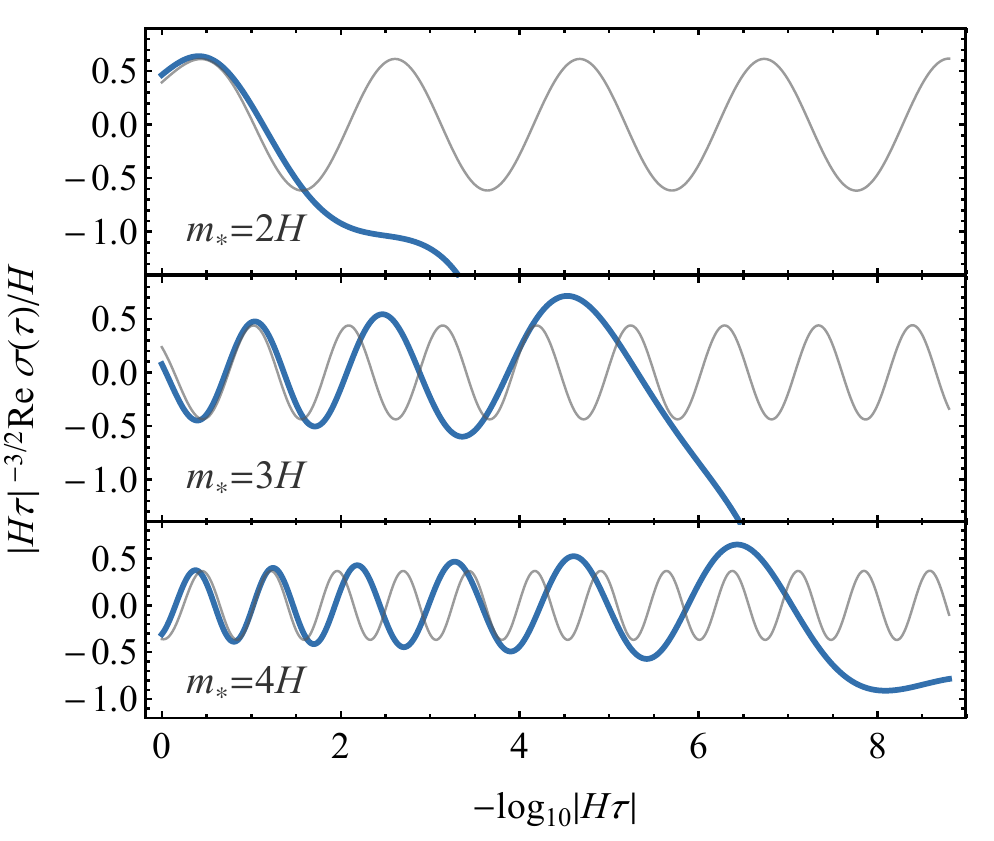}
  \caption{The real part of the mode functions with time dependent masses (\ref{msigma}) (dark blue). In all plots we take $\al=1$ and $\ep_*=0.02$. For comparison, the corresponding constant-mass mode functions are shown in gray curves.}
  \label{fig_mode}
\end{figure}

With a rolling background $\phi_0(t)\simeq\phi_{0*}-|\dot\phi_{0*}|(t-t_*)=\sqrt{2\ep}\Mp H_*(t-t_*)$ (we assume $\dot\phi_0<0$ without loss of generality), the $\phi$-dependence gets translated to a time dependence:
\begin{align}
\label{msigma}
  m_\text{eff}(t)=m_{*}e^{-\al\sqrt{2\ep}H t},
\end{align}
where $t$ is the physical time. Here and below we remove the $*$ in $\ep$ and $H$. Decomposing the $\si$ field into Fourier modes with fixed 3d momentum $\mb k$, we can find the equation of motion for a single mode $\si_{\mb k}$ as
\begin{align}
\label{sigmaeq}
  \si_{\mb k}'' -\FR{2}{\tau}\si_{\mb k}' +\big(k^2+m_{*}^2|H\tau|^{2\al\sqrt{2\ep}-2}\big)\si_{\mb k} =0,
\end{align}
where a prime denotes the derivative with respect to conformal time $\tau$, which is related to $t$ via $e^{Ht}=a=-1/(H\tau)$. With the usual Bunch-Davies initial condition for the mode, the above equation has a unique solution up to an irrelevant overall phase. 

We are not aware of any named special functions that solve this equation, but it is possible to proceed analytically by making approximations. For example, when the mode leaves the horizon at late times ($|k\tau|\ll 1$) and when the mass $m(\tau)$ is still much greater than Hubble, we can apply the WKB approximation. Rewriting $\si_{\mb k}(\tau)=\tau\chi_{\mb k}(\tau)$, the equation for $\chi_k$ is
\begin{align}
\label{chieq}
  \chi_{\mb k}'' +\big(k^2+m_{*}^2|H\tau|^{2\al\sqrt{2\ep}-2}-2\tau^{-2}\big)\chi_{\mb k} =0.
\end{align}
At late times, we can neglect the momentum, and the WKB solution can be found to be
\begin{align}
\label{eq_WKB}
  \chi_{\mb k}(\tau)\simeq &~ A e^{+\ii \vartheta(\tau)}+B e^{-\ii \vartheta(\tau)}.
\end{align}
When the effective mass $m_\text{eff}=m_{*}|H\tau|^{\al\sqrt{2\ep}} \lesssim H$, the WKB solution stops oscillating and becomes invalid.  Therefore the above approximation works only when $m_\text{eff}\gg H$, in which case the phase $\vartheta(\tau)$ can be approximated by: 
\begin{align}
  \vartheta(\tau)&\simeq  \FR{1}{\al\sqrt{2\ep}}\FR{m_{*}}{H}\big(|H\tau|^{\al\sqrt{2\ep}}-1\big)\n\\
  &=\FR{m_{*}}{H}\log|H\tau|\big(1+\al\sqrt{\fr{\ep}{2}}\log|H\tau|+\order{\ep}\big).
\end{align}
Therefore, the oscillation frequency of the mode function in the late time limit is itself time dependent, just as expected. On the other hand, it is possible to solve the mode equation (\ref{sigmaeq}) numerically without taking the WKB approximation. We show the solutions with several choices of parameters in Fig.\;\ref{fig_mode}, in which the mild time dependence of the oscillation frequency is evident.

\noindent{\bf Power spectrum in large-field models.} It is important to check the scale dependence of the power spectrum of the inflaton fluctuation $\varphi$ induced by the $\si$-$\varphi$ coupling. This coupling arises from the $\lam_5$ term in (\ref{eq_ActionTD}) evaluated with the inflation background:
\bge
  \Delta\ld=a^3\mu \varphi'\si,
\ede
where $\mu\equiv-\lam_5\dot\phi_0$. A full calculation of the power spectrum should include slow-roll corrections everywhere consistently, including the coefficients of $\si_{\mb k}'$ and $\varphi_{\mb k}'$ in the mode equation, together with the slow-roll induced mass correction to $\si_{\mb k}$. However, in order to isolate the effect of a time-dependent mass $m_\text{eff}$, we will only retain the time dependence in $m_\text{eff}$, and assume a dS limit for the background evolution. 
The resulting power spectrum then reflects the scale dependence from $m_\text{eff}(\tau)$. We will check whether this scale dependence contains any features (wiggles), and if not, whether the monotonic scale dependence would be too large to be compatible with current observations.

To calculate the power spectrum analytically is difficult, because the mass of $\si$ is time dependent, and also because the mixing between $\si$ and $\varphi$ could be strong, which invalidates a perturbative treatment. Therefore, we numerically evolve the following set of equations, with the usual Bunch-Davies initial condition.
\begin{align}
\label{eq_EoM}
  &\si_\mb{k}'' -\FR{2}{\tau}\si_\mb{k}' +\big(k^2+m_{*}^2|H\tau|^{2\al\sqrt{2\ep}-2}\big)\si_\mb{k} =\FR{-\mu}{H\tau}\varphi_\mb{k},\\
  &\varphi_\mb{k}'' -\FR{2}{\tau}\varphi_\mb{k}' +k^2\varphi_\mb{k} =-\FR{3\mu}{H\tau^2}\si_\mb{k}+\FR{\mu}{H\tau}\si_\mb{k}'.
\end{align}
The power spectrum $P_\zeta$ is then computed by
\begin{align}
\label{eq_PS}
  P_\zeta(k)=(H/\dot\phi_0)^2(k^3/2\pi^2)\big\la\varphi_{-\mb k}(\tau_f)\varphi_{\mb k}(\tau_f)\big\ra',
\end{align}
where the prime $\la\cdots\ra'$ means the momentum-conserving $\de$-function is removed, and $\tau_f$ is chosen so that $|k\tau_f|\ll 1$ for all relevant $k$. This is effectively computing the following set of diagrams:
\bge
\label{fd_PS}
\parbox{0.4\textwidth}{\includegraphics[width=0.4\textwidth]{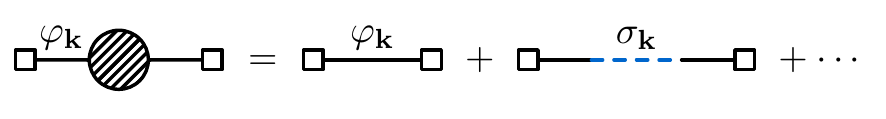}}.
\ede

\begin{figure}[t]
  \centering
  \includegraphics[width=0.41\textwidth]{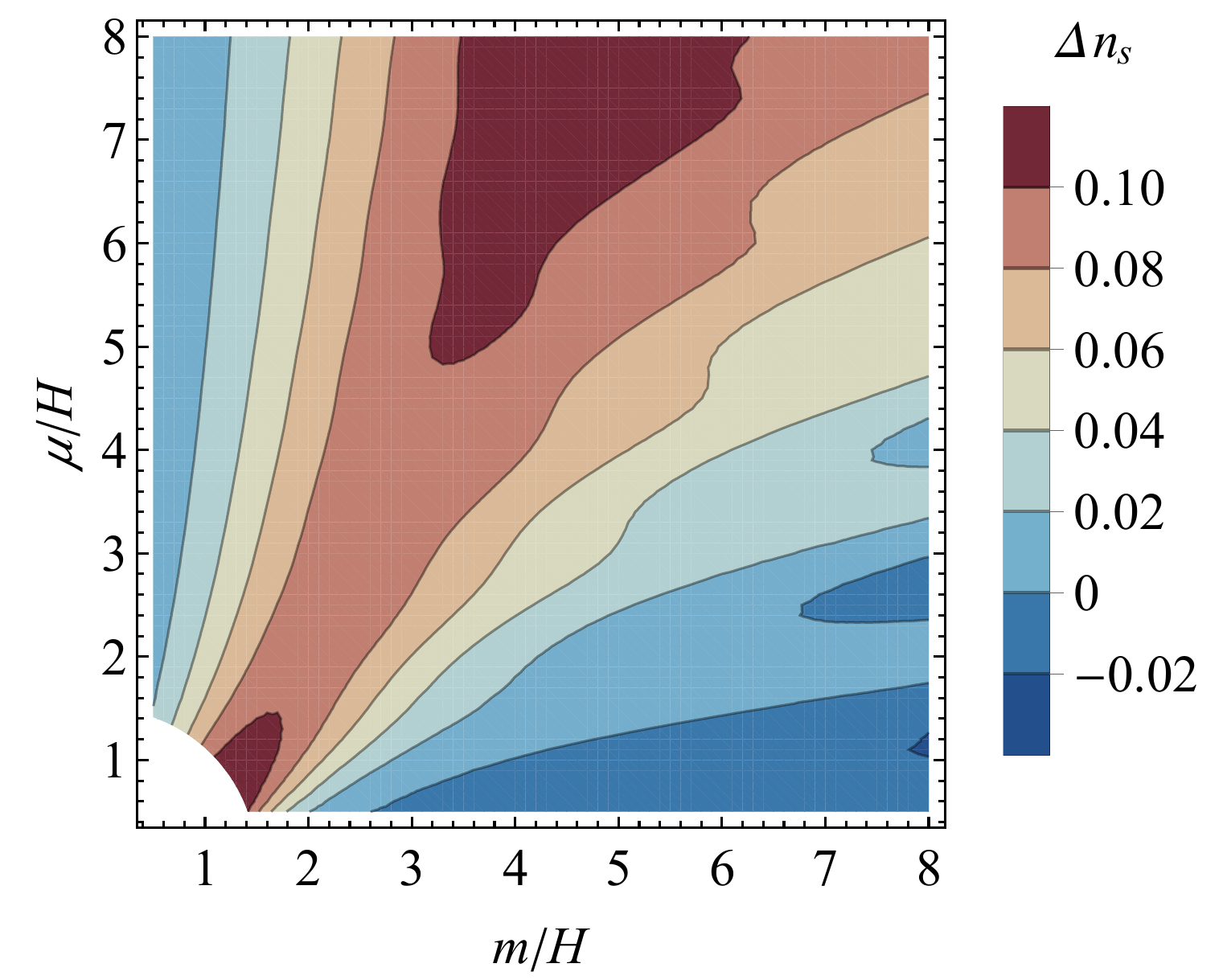}
  \caption{The correction to scalar tilt, $\Delta n_s$, from an intermediate state with time-dependent mass given in (\ref{msigma}), plotted on the plane of $\si$ mass $m$ at a reference scale and the mixing parameter $\mu$ between $\si$ and $\de\phi$. In this plot we fix $\ep=1.25\times 10^{-3}$, which corresponds to $r=0.02$ ($<0.02$) when $\mu=0$ $(>0)$. The lower-left corner with $\sqrt{m^2+\mu^2}<3H/2$, where no oscillation signals occur,  is excluded. }
  \label{fig_ns}
\end{figure}

In our numerical result we do not observe any oscillatory signal, but only a smooth scale dependence, as expected. So, the scale dependence introduced by the time-dependent mass is degenerate with the scale dependence at zeroth order (i.e., the scale dependence from the slow-roll potential). For this reason, we do not consider it an independent observable. But we do need to check that the new scale dependence from the heavy field mixing is not much larger than the observed value. Otherwise we would need significant tuning between the background contribution and the massive-state correction. The scalar tilt $ n_s-1 \equiv \di\log P_\zeta/\di \log k$ induced by the time-dependent mass is shown in Fig.\ \ref{fig_ns} on the $\mu$-$m$ plane. This result is to be added with the slow-roll contributions, like those considered in \cite{An:2017hlx}, to get the observed value from CMB measurement $n_s^\text{(CMB)}\simeq 0.967$ \cite{Planck:2018jri}. We see that the correction $\Delta n_s$ in Fig.\ \ref{fig_ns} stays at the same order as the slow-roll contribution $n_s^\text{(CMB)}-1\simeq - 0.033$ for most parameter space. 
Therefore we conclude that the scale dependence from the time-dependent mass is compatible with the current observations for most of parameter space we are interested in. In the appendix we elaborate on our calculation of $n_s$ and its relation to previous works.

\noindent{\bf The squeezed bispectrum.} As shown above,  large-field inflation can introduce a relatively large scale dependence to the mass of a spectator field $\si$. Now we study the consequence of this scale dependence in the 3-point correlator mediated by $\si$. This is a well studied ``discovery channel'' for the cosmological collider, and it was found that the largest signal comes from the following diagram, where the blobs again include arbitrary numbers of two-point mixing insertions \cite{Chen:2017ryl,Wang:2019gbi}. 
\bge
\label{eq_3ptdiagram}
\parbox{0.23\textwidth}{\includegraphics[width=0.23\textwidth]{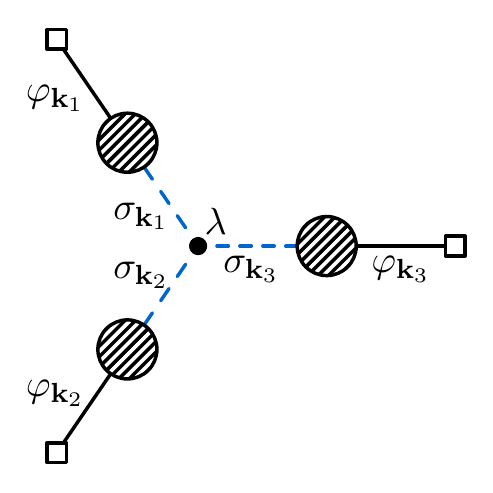}}
\ede
The diagram (\ref{eq_3ptdiagram}) can be computed in the following way
\begin{align}
\label{eq_bispect}
\la \varphi_{\mb k_1}\varphi_{\mb k_2}\varphi_{\mb k_3}\ra'=&\; 2\lam_3\,\text{Im}\int_{-\infty}^0 \di\tau\, a^4 \prod_{i=1}^3\la \si_{\mb k_i}(\tau)\varphi_{-\mb k_i}(\tau_f)\ra,
\end{align}
from which we can get the shape function $\mathcal{S}$ by
\begin{align}
\label{eq_shape}
\mathcal{S}(k_1,k_2,k_3)=&\;\FR{(k_1k_2k_3)^2}{2\pi P_\zeta^{1/2}H^3}\la\varphi_{\mb k_1}\varphi_{\mb k_2}\varphi_{\mb k_3}\ra'.
\end{align}
Again, it is virtually impossible to find a closed analytical result for this amplitude. We will therefore first try to understand the qualitative features of this process by taking analytical approximations, and then present full numerical results.

\begin{figure}
  \centering
  \includegraphics[width=0.48\textwidth]{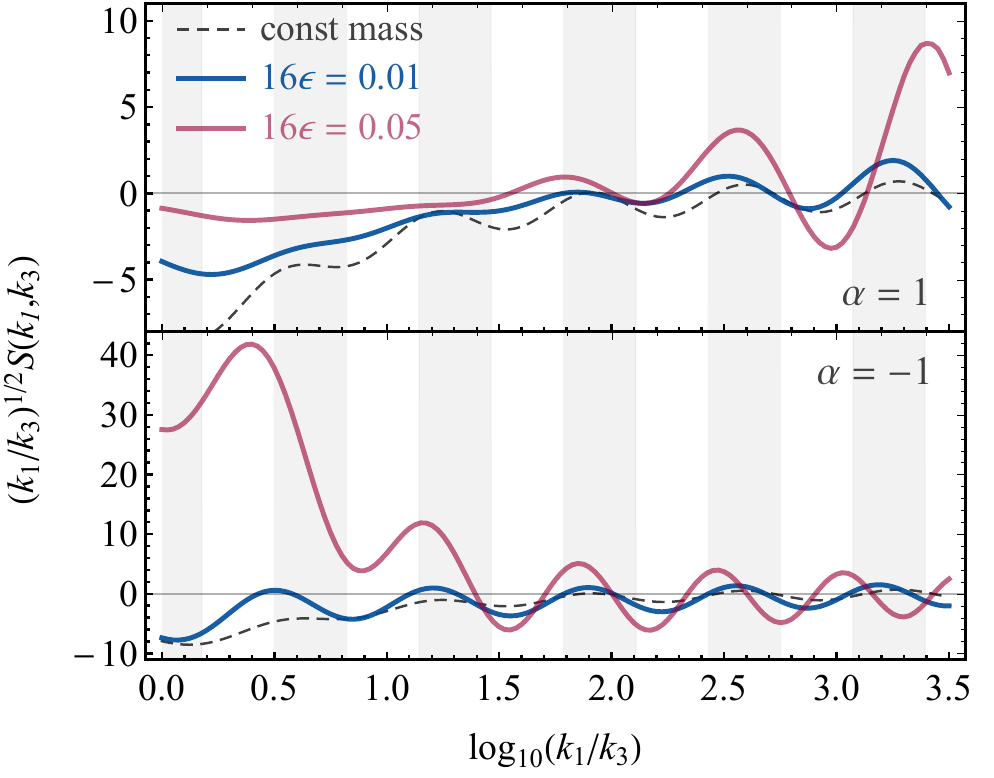}
  \caption{The shape function (\ref{eq_shape}) from the process (\ref{eq_3ptdiagram}). In both panels we take $m_*=\mu=3H$ and $\lam_3=H$. The gray shades indicate the expected frequency $\nu_\mu=\sqrt{m_*^2+\mu^2}$ for constant mass in the squeezed limit. The blue and red curves show the signals from time-dependent masses, in which both the amplitude and the frequency of the oscillations change visibly with $k_1/k_3$.}
  \label{fig_signal}
\end{figure}

For a crude analytical understanding, we focus on the oscillatory signal in the squeezed limit $k_3\ll k_1\simeq k_2$ and consider the perturbative regime $\mu\ll m$ and $m/H\gg 1$. The signal with constant mass can be estimated as \cite{Chen:2015lza,Wang:2019gbi}
\begin{align}
\label{eq_signalsize}
  \mathcal{S}_\text{signal} \simeq \FR{\lam_3\mu^3}{\nu^{5/2}H^4}e^{-\pi\nu}\sqrt{\FR{k_3}{k_1}}\sin\Big(\nu\log\FR{k_3}{k_1}+\vartheta\Big),
\end{align}
where $\nu\equiv\sqrt{m^2/H^2-9/4}$. We emphasize that, for constant $m$, the signal has a fixed oscillation frequency $\nu$ and its size has a fixed scaling with the momentum ratio, characterized by the power law $(k_3/k_1)^{1/2}$.

For our case where the mass has a weak time dependence, we can still use (\ref{eq_signalsize}), but this time evaluate $m=m(\tau)$ at the saddle point $\tau\simeq -\nu/(2k_1)$ of the integrand $\sim e^{\pm 2\ii k_1\tau}(-\tau)^{\pm\ii\nu}$ where the signal receives most contribution from the integral (\ref{eq_bispect}). Consequently, the time dependence of $m$ is translated to a scale dependence of $\nu$ in (\ref{eq_signalsize}), which appears in both the signal frequency and in the overall size, and this immediately leads to one of the main results of this work: The cosmological collider signal from a time-dependent mass has scale-dependent oscillation frequency, and its overall size scales with $k_3/k_1$ differently from the power law $(k_3/k_1)^{1/2}$ for a constant mass. The deviation in the scaling with $k_3/k_1$ is quite significant due to the exponential factor $e^{-\pi\nu}$, which makes the signal depend very sensitively on the mass $m$.

In the strongly coupled regime ($\mu\gtrsim m$), no closed analytical estimate like (\ref{eq_signalsize}) is known. However, it is known that the oscillation frequency of the signal is modified to $\nu\to \nu_\mu\equiv\sqrt{(m^2+\mu^2)/H^2-9/4}$~\cite{An:2017hlx}. We can also anticipate that the signal size has a sensitive dependence on the mass $m$ so long as $\mu$ is not too much greater than $m$. Therefore, our main result, namely, the scale-dependent oscillation frequency and the signal size, still holds. 

The above analytical arguments rely on approximations in several limits which are never really reached by realistic parameters. Therefore a direct numerical approach is indispensable to get the signal shape precisely. We compute (\ref{eq_bispect}) numerically and get the full bispectrum, including both the background and the signal. Several examples are shown in Fig.\;\ref{fig_signal}. We show the signals with the mass decreasing/increasing in the upper/lower panel, respectively. Compared with the bispectrum of constant mass (dashed curves), the slow changes of the frequency and the amplitude are evident in both cases. 

\noindent{\bf Discussions.} Large-field inflation models typically show significant scale dependence that can distort the cosmological collider signals. In this Letter we show that a direct coupling between a massive state and the inflaton can lead to an oscillatory signal in the bispectrum, with significant and nonstandard momentum-ratio dependence in both the size and the frequency. We use quasi-single-field inflation and an exponential ansatz for the mass to illustrate the point, but we stress that the signal distortion found in this work depends mainly on the large field excursion and the direct coupling. Thus we expect that similar effects should also show up in a broader class of models with these ingredients. 

Phenomenologically, our work shows that the momentum-ratio dependence of the signal size and frequency can be informative probes of rich dynamics during inflation. (See \cite{Lu:2021wxu} for a related example.) Therefore, our work invites efforts on developing more realistic templates allowing for the possibility that the signal size and the frequency could be momentum-ratio dependent.

There are other sources of scale dependence than the one considered in this Letter, including the slow-roll background. These have been explored in the context of primordial standard and nonstandard clocks \cite{Chen:2015lza,Chen:2018cgg,Wang:2020aqc}. We expect the effect to be weaker than ours in large-field inflation, as detailed before. There is also a known slow change of signal frequency in the not-so-squeezed configurations. This change is automatically included in a numerical approach, and can also be resolved analytically by pushing the calculation to higher orders in the momentum ratio. 

In this Letter we only considered the effect of time-dependent mass. More generally, it would be interesting to explore the similar time dependences in other parameters. In particular, a time-dependent two-point mixing parameter $\mu$ could induce a similar change in the signal frequency. It would also be interesting to incorporate these nonstandard momentum-ratio dependences in template-based Fisher forecasts for future observations. We leave these topics for future studies.

\begin{acknowledgments}
\noindent\textbf{Acknowledgements.}
MR is supported in part by the NASA Grant 80NSSC20K0506 and the DOE Grant DE-SC0013607.
 LTW is supported by the DOE grant DE-SC0013642. 
ZZX is supported in part by the National Key R\&D Program of China (2021YFC2203100), an Open Research Fund of the Key Laboratory of Particle Astrophysics and Cosmology, Ministry of Education of China, and a Tsinghua University Initiative Scientific Research Program.
\end{acknowledgments}

\bibliography{CosmoCollider} 

\begin{thebibliography}{38}%
\makeatletter
\providecommand \@ifxundefined [1]{%
 \@ifx{#1\undefined}
}%
\providecommand \@ifnum [1]{%
 \ifnum #1\expandafter \@firstoftwo
 \else \expandafter \@secondoftwo
 \fi
}%
\providecommand \@ifx [1]{%
 \ifx #1\expandafter \@firstoftwo
 \else \expandafter \@secondoftwo
 \fi
}%
\providecommand \natexlab [1]{#1}%
\providecommand \enquote  [1]{``#1''}%
\providecommand \bibnamefont  [1]{#1}%
\providecommand \bibfnamefont [1]{#1}%
\providecommand \citenamefont [1]{#1}%
\providecommand \href@noop [0]{\@secondoftwo}%
\providecommand \href [0]{\begingroup \@sanitize@url \@href}%
\providecommand \@href[1]{\@@startlink{#1}\@@href}%
\providecommand \@@href[1]{\endgroup#1\@@endlink}%
\providecommand \@sanitize@url [0]{\catcode `\\12\catcode `\$12\catcode
  `\&12\catcode `\#12\catcode `\^12\catcode `\_12\catcode `\%12\relax}%
\providecommand \@@startlink[1]{}%
\providecommand \@@endlink[0]{}%
\providecommand \url  [0]{\begingroup\@sanitize@url \@url }%
\providecommand \@url [1]{\endgroup\@href {#1}{\urlprefix }}%
\providecommand \urlprefix  [0]{URL }%
\providecommand \Eprint [0]{\href }%
\providecommand \doibase [0]{http://dx.doi.org/}%
\providecommand \selectlanguage [0]{\@gobble}%
\providecommand \bibinfo  [0]{\@secondoftwo}%
\providecommand \bibfield  [0]{\@secondoftwo}%
\providecommand \translation [1]{[#1]}%
\providecommand \BibitemOpen [0]{}%
\providecommand \bibitemStop [0]{}%
\providecommand \bibitemNoStop [0]{.\EOS\space}%
\providecommand \EOS [0]{\spacefactor3000\relax}%
\providecommand \BibitemShut  [1]{\csname bibitem#1\endcsname}%
\let\auto@bib@innerbib\@empty
\bibitem [{\citenamefont {Akrami}\ \emph {et~al.}(2020)\citenamefont {Akrami}
  \emph {et~al.}}]{Planck:2018jri}%
  \BibitemOpen
  \bibfield  {author} {\bibinfo {author} {\bibfnamefont {Y.}~\bibnamefont
  {Akrami}} \emph {et~al.} (\bibinfo {collaboration} {Planck}),\ }\href
  {\doibase 10.1051/0004-6361/201833887} {\bibfield  {journal} {\bibinfo
  {journal} {Astron. Astrophys.}\ }\textbf {\bibinfo {volume} {641}},\ \bibinfo
  {pages} {A10} (\bibinfo {year} {2020})},\ \Eprint
  {http://arxiv.org/abs/1807.06211} {arXiv:1807.06211 [astro-ph.CO]}
  \BibitemShut {NoStop}%
\bibitem [{\citenamefont {Lyth}(1997)}]{Lyth:1996im}%
  \BibitemOpen
  \bibfield  {author} {\bibinfo {author} {\bibfnamefont {D.~H.}\ \bibnamefont
  {Lyth}},\ }\href {\doibase 10.1103/PhysRevLett.78.1861} {\bibfield  {journal}
  {\bibinfo  {journal} {Phys. Rev. Lett.}\ }\textbf {\bibinfo {volume} {78}},\
  \bibinfo {pages} {1861} (\bibinfo {year} {1997})},\ \Eprint
  {http://arxiv.org/abs/hep-ph/9606387} {arXiv:hep-ph/9606387} \BibitemShut
  {NoStop}%
\bibitem [{\citenamefont {Baumann}\ and\ \citenamefont
  {Green}(2012)}]{Baumann:2011ws}%
  \BibitemOpen
  \bibfield  {author} {\bibinfo {author} {\bibfnamefont {D.}~\bibnamefont
  {Baumann}}\ and\ \bibinfo {author} {\bibfnamefont {D.}~\bibnamefont
  {Green}},\ }\href {\doibase 10.1088/1475-7516/2012/05/017} {\bibfield
  {journal} {\bibinfo  {journal} {JCAP}\ }\textbf {\bibinfo {volume} {05}},\
  \bibinfo {pages} {017} (\bibinfo {year} {2012})},\ \Eprint
  {http://arxiv.org/abs/1111.3040} {arXiv:1111.3040 [hep-th]} \BibitemShut
  {NoStop}%
\bibitem [{\citenamefont {Mirbabayi}\ \emph {et~al.}(2015)\citenamefont
  {Mirbabayi}, \citenamefont {Senatore}, \citenamefont {Silverstein},\ and\
  \citenamefont {Zaldarriaga}}]{Mirbabayi:2014jqa}%
  \BibitemOpen
  \bibfield  {author} {\bibinfo {author} {\bibfnamefont {M.}~\bibnamefont
  {Mirbabayi}}, \bibinfo {author} {\bibfnamefont {L.}~\bibnamefont {Senatore}},
  \bibinfo {author} {\bibfnamefont {E.}~\bibnamefont {Silverstein}}, \ and\
  \bibinfo {author} {\bibfnamefont {M.}~\bibnamefont {Zaldarriaga}},\ }\href
  {\doibase 10.1103/PhysRevD.91.063518} {\bibfield  {journal} {\bibinfo
  {journal} {Phys. Rev. D}\ }\textbf {\bibinfo {volume} {91}},\ \bibinfo
  {pages} {063518} (\bibinfo {year} {2015})},\ \Eprint
  {http://arxiv.org/abs/1412.0665} {arXiv:1412.0665 [hep-th]} \BibitemShut
  {NoStop}%
\bibitem [{\citenamefont {Ooguri}\ and\ \citenamefont
  {Vafa}(2007)}]{Ooguri:2006in}%
  \BibitemOpen
  \bibfield  {author} {\bibinfo {author} {\bibfnamefont {H.}~\bibnamefont
  {Ooguri}}\ and\ \bibinfo {author} {\bibfnamefont {C.}~\bibnamefont {Vafa}},\
  }\href {\doibase 10.1016/j.nuclphysb.2006.10.033} {\bibfield  {journal}
  {\bibinfo  {journal} {Nucl. Phys. B}\ }\textbf {\bibinfo {volume} {766}},\
  \bibinfo {pages} {21} (\bibinfo {year} {2007})},\ \Eprint
  {http://arxiv.org/abs/hep-th/0605264} {arXiv:hep-th/0605264} \BibitemShut
  {NoStop}%
\bibitem [{\citenamefont {Klaewer}\ and\ \citenamefont
  {Palti}(2017)}]{Klaewer:2016kiy}%
  \BibitemOpen
  \bibfield  {author} {\bibinfo {author} {\bibfnamefont {D.}~\bibnamefont
  {Klaewer}}\ and\ \bibinfo {author} {\bibfnamefont {E.}~\bibnamefont
  {Palti}},\ }\href {\doibase 10.1007/JHEP01(2017)088} {\bibfield  {journal}
  {\bibinfo  {journal} {JHEP}\ }\textbf {\bibinfo {volume} {01}},\ \bibinfo
  {pages} {088} (\bibinfo {year} {2017})},\ \Eprint
  {http://arxiv.org/abs/1610.00010} {arXiv:1610.00010 [hep-th]} \BibitemShut
  {NoStop}%
\bibitem [{\citenamefont {Heidenreich}\ \emph {et~al.}(2018)\citenamefont
  {Heidenreich}, \citenamefont {Reece},\ and\ \citenamefont
  {Rudelius}}]{Heidenreich:2018kpg}%
  \BibitemOpen
  \bibfield  {author} {\bibinfo {author} {\bibfnamefont {B.}~\bibnamefont
  {Heidenreich}}, \bibinfo {author} {\bibfnamefont {M.}~\bibnamefont {Reece}},
  \ and\ \bibinfo {author} {\bibfnamefont {T.}~\bibnamefont {Rudelius}},\
  }\href {\doibase 10.1103/PhysRevLett.121.051601} {\bibfield  {journal}
  {\bibinfo  {journal} {Phys. Rev. Lett.}\ }\textbf {\bibinfo {volume} {121}},\
  \bibinfo {pages} {051601} (\bibinfo {year} {2018})},\ \Eprint
  {http://arxiv.org/abs/1802.08698} {arXiv:1802.08698 [hep-th]} \BibitemShut
  {NoStop}%
\bibitem [{\citenamefont {Grimm}\ \emph {et~al.}(2018)\citenamefont {Grimm},
  \citenamefont {Palti},\ and\ \citenamefont {Valenzuela}}]{Grimm:2018ohb}%
  \BibitemOpen
  \bibfield  {author} {\bibinfo {author} {\bibfnamefont {T.~W.}\ \bibnamefont
  {Grimm}}, \bibinfo {author} {\bibfnamefont {E.}~\bibnamefont {Palti}}, \ and\
  \bibinfo {author} {\bibfnamefont {I.}~\bibnamefont {Valenzuela}},\ }\href
  {\doibase 10.1007/JHEP08(2018)143} {\bibfield  {journal} {\bibinfo  {journal}
  {JHEP}\ }\textbf {\bibinfo {volume} {08}},\ \bibinfo {pages} {143} (\bibinfo
  {year} {2018})},\ \Eprint {http://arxiv.org/abs/1802.08264} {arXiv:1802.08264
  [hep-th]} \BibitemShut {NoStop}%
\bibitem [{\citenamefont {Dine}\ and\ \citenamefont
  {Seiberg}(1985)}]{Dine:1985he}%
  \BibitemOpen
  \bibfield  {author} {\bibinfo {author} {\bibfnamefont {M.}~\bibnamefont
  {Dine}}\ and\ \bibinfo {author} {\bibfnamefont {N.}~\bibnamefont {Seiberg}},\
  }\href {\doibase 10.1016/0370-2693(85)90927-X} {\bibfield  {journal}
  {\bibinfo  {journal} {Phys. Lett. B}\ }\textbf {\bibinfo {volume} {162}},\
  \bibinfo {pages} {299} (\bibinfo {year} {1985})}\BibitemShut {NoStop}%
\bibitem [{\citenamefont {Scalisi}\ and\ \citenamefont
  {Valenzuela}(2019)}]{Scalisi:2018eaz}%
  \BibitemOpen
  \bibfield  {author} {\bibinfo {author} {\bibfnamefont {M.}~\bibnamefont
  {Scalisi}}\ and\ \bibinfo {author} {\bibfnamefont {I.}~\bibnamefont
  {Valenzuela}},\ }\href {\doibase 10.1007/JHEP08(2019)160} {\bibfield
  {journal} {\bibinfo  {journal} {JHEP}\ }\textbf {\bibinfo {volume} {08}},\
  \bibinfo {pages} {160} (\bibinfo {year} {2019})},\ \Eprint
  {http://arxiv.org/abs/1812.07558} {arXiv:1812.07558 [hep-th]} \BibitemShut
  {NoStop}%
\bibitem [{\citenamefont {Silverstein}\ and\ \citenamefont
  {Westphal}(2008)}]{Silverstein:2008sg}%
  \BibitemOpen
  \bibfield  {author} {\bibinfo {author} {\bibfnamefont {E.}~\bibnamefont
  {Silverstein}}\ and\ \bibinfo {author} {\bibfnamefont {A.}~\bibnamefont
  {Westphal}},\ }\href {\doibase 10.1103/PhysRevD.78.106003} {\bibfield
  {journal} {\bibinfo  {journal} {Phys. Rev.}\ }\textbf {\bibinfo {volume}
  {D78}},\ \bibinfo {pages} {106003} (\bibinfo {year} {2008})},\ \Eprint
  {http://arxiv.org/abs/0803.3085} {arXiv:0803.3085 [hep-th]} \BibitemShut
  {NoStop}%
\bibitem [{\citenamefont {McAllister}\ \emph {et~al.}(2010)\citenamefont
  {McAllister}, \citenamefont {Silverstein},\ and\ \citenamefont
  {Westphal}}]{McAllister:2008hb}%
  \BibitemOpen
  \bibfield  {author} {\bibinfo {author} {\bibfnamefont {L.}~\bibnamefont
  {McAllister}}, \bibinfo {author} {\bibfnamefont {E.}~\bibnamefont
  {Silverstein}}, \ and\ \bibinfo {author} {\bibfnamefont {A.}~\bibnamefont
  {Westphal}},\ }\href {\doibase 10.1103/PhysRevD.82.046003} {\bibfield
  {journal} {\bibinfo  {journal} {Phys. Rev.}\ }\textbf {\bibinfo {volume}
  {D82}},\ \bibinfo {pages} {046003} (\bibinfo {year} {2010})},\ \Eprint
  {http://arxiv.org/abs/0808.0706} {arXiv:0808.0706 [hep-th]} \BibitemShut
  {NoStop}%
\bibitem [{\citenamefont {McAllister}\ \emph {et~al.}(2014)\citenamefont
  {McAllister}, \citenamefont {Silverstein}, \citenamefont {Westphal},\ and\
  \citenamefont {Wrase}}]{McAllister:2014mpa}%
  \BibitemOpen
  \bibfield  {author} {\bibinfo {author} {\bibfnamefont {L.}~\bibnamefont
  {McAllister}}, \bibinfo {author} {\bibfnamefont {E.}~\bibnamefont
  {Silverstein}}, \bibinfo {author} {\bibfnamefont {A.}~\bibnamefont
  {Westphal}}, \ and\ \bibinfo {author} {\bibfnamefont {T.}~\bibnamefont
  {Wrase}},\ }\href {\doibase 10.1007/JHEP09(2014)123} {\bibfield  {journal}
  {\bibinfo  {journal} {JHEP}\ }\textbf {\bibinfo {volume} {09}},\ \bibinfo
  {pages} {123} (\bibinfo {year} {2014})},\ \Eprint
  {http://arxiv.org/abs/1405.3652} {arXiv:1405.3652 [hep-th]} \BibitemShut
  {NoStop}%
\bibitem [{\citenamefont {Dong}\ \emph {et~al.}(2011)\citenamefont {Dong},
  \citenamefont {Horn}, \citenamefont {Silverstein},\ and\ \citenamefont
  {Westphal}}]{Dong:2010in}%
  \BibitemOpen
  \bibfield  {author} {\bibinfo {author} {\bibfnamefont {X.}~\bibnamefont
  {Dong}}, \bibinfo {author} {\bibfnamefont {B.}~\bibnamefont {Horn}}, \bibinfo
  {author} {\bibfnamefont {E.}~\bibnamefont {Silverstein}}, \ and\ \bibinfo
  {author} {\bibfnamefont {A.}~\bibnamefont {Westphal}},\ }\href {\doibase
  10.1103/PhysRevD.84.026011} {\bibfield  {journal} {\bibinfo  {journal} {Phys.
  Rev. D}\ }\textbf {\bibinfo {volume} {84}},\ \bibinfo {pages} {026011}
  (\bibinfo {year} {2011})},\ \Eprint {http://arxiv.org/abs/1011.4521}
  {arXiv:1011.4521 [hep-th]} \BibitemShut {NoStop}%
\bibitem [{\citenamefont {Flauger}\ \emph {et~al.}(2017)\citenamefont
  {Flauger}, \citenamefont {Mirbabayi}, \citenamefont {Senatore},\ and\
  \citenamefont {Silverstein}}]{Flauger:2016idt}%
  \BibitemOpen
  \bibfield  {author} {\bibinfo {author} {\bibfnamefont {R.}~\bibnamefont
  {Flauger}}, \bibinfo {author} {\bibfnamefont {M.}~\bibnamefont {Mirbabayi}},
  \bibinfo {author} {\bibfnamefont {L.}~\bibnamefont {Senatore}}, \ and\
  \bibinfo {author} {\bibfnamefont {E.}~\bibnamefont {Silverstein}},\ }\href
  {\doibase 10.1088/1475-7516/2017/10/058} {\bibfield  {journal} {\bibinfo
  {journal} {JCAP}\ }\textbf {\bibinfo {volume} {1710}},\ \bibinfo {pages}
  {058} (\bibinfo {year} {2017})},\ \Eprint {http://arxiv.org/abs/1606.00513}
  {arXiv:1606.00513 [hep-th]} \BibitemShut {NoStop}%
\bibitem [{\citenamefont {An}\ \emph {et~al.}(2020)\citenamefont {An},
  \citenamefont {Lyu}, \citenamefont {Wang},\ and\ \citenamefont
  {Zhou}}]{An:2020fff}%
  \BibitemOpen
  \bibfield  {author} {\bibinfo {author} {\bibfnamefont {H.}~\bibnamefont
  {An}}, \bibinfo {author} {\bibfnamefont {K.-F.}\ \bibnamefont {Lyu}},
  \bibinfo {author} {\bibfnamefont {L.-T.}\ \bibnamefont {Wang}}, \ and\
  \bibinfo {author} {\bibfnamefont {S.}~\bibnamefont {Zhou}},\ }\href@noop {}
  {\  (\bibinfo {year} {2020})},\ \Eprint {http://arxiv.org/abs/2009.12381}
  {arXiv:2009.12381 [astro-ph.CO]} \BibitemShut {NoStop}%
\bibitem [{\citenamefont {An}\ \emph {et~al.}(2022)\citenamefont {An},
  \citenamefont {Lyu}, \citenamefont {Wang},\ and\ \citenamefont
  {Zhou}}]{An:2022cce}%
  \BibitemOpen
  \bibfield  {author} {\bibinfo {author} {\bibfnamefont {H.}~\bibnamefont
  {An}}, \bibinfo {author} {\bibfnamefont {K.-F.}\ \bibnamefont {Lyu}},
  \bibinfo {author} {\bibfnamefont {L.-T.}\ \bibnamefont {Wang}}, \ and\
  \bibinfo {author} {\bibfnamefont {S.}~\bibnamefont {Zhou}},\ }\href@noop {}
  {\  (\bibinfo {year} {2022})},\ \Eprint {http://arxiv.org/abs/2201.05171}
  {arXiv:2201.05171 [astro-ph.CO]} \BibitemShut {NoStop}%
\bibitem [{\citenamefont {Chen}\ and\ \citenamefont
  {Wang}(2010)}]{Chen:2009zp}%
  \BibitemOpen
  \bibfield  {author} {\bibinfo {author} {\bibfnamefont {X.}~\bibnamefont
  {Chen}}\ and\ \bibinfo {author} {\bibfnamefont {Y.}~\bibnamefont {Wang}},\
  }\href {\doibase 10.1088/1475-7516/2010/04/027} {\bibfield  {journal}
  {\bibinfo  {journal} {JCAP}\ }\textbf {\bibinfo {volume} {1004}},\ \bibinfo
  {pages} {027} (\bibinfo {year} {2010})},\ \Eprint
  {http://arxiv.org/abs/0911.3380} {arXiv:0911.3380 [hep-th]} \BibitemShut
  {NoStop}%
\bibitem [{\citenamefont {Noumi}\ \emph {et~al.}(2013)\citenamefont {Noumi},
  \citenamefont {Yamaguchi},\ and\ \citenamefont {Yokoyama}}]{Noumi:2012vr}%
  \BibitemOpen
  \bibfield  {author} {\bibinfo {author} {\bibfnamefont {T.}~\bibnamefont
  {Noumi}}, \bibinfo {author} {\bibfnamefont {M.}~\bibnamefont {Yamaguchi}}, \
  and\ \bibinfo {author} {\bibfnamefont {D.}~\bibnamefont {Yokoyama}},\ }\href
  {\doibase 10.1007/JHEP06(2013)051} {\bibfield  {journal} {\bibinfo  {journal}
  {JHEP}\ }\textbf {\bibinfo {volume} {06}},\ \bibinfo {pages} {051} (\bibinfo
  {year} {2013})},\ \Eprint {http://arxiv.org/abs/1211.1624} {arXiv:1211.1624
  [hep-th]} \BibitemShut {NoStop}%
\bibitem [{\citenamefont {Arkani-Hamed}\ and\ \citenamefont
  {Maldacena}(2015)}]{Arkani-Hamed:2015bza}%
  \BibitemOpen
  \bibfield  {author} {\bibinfo {author} {\bibfnamefont {N.}~\bibnamefont
  {Arkani-Hamed}}\ and\ \bibinfo {author} {\bibfnamefont {J.}~\bibnamefont
  {Maldacena}},\ }\href@noop {} {\  (\bibinfo {year} {2015})},\ \Eprint
  {http://arxiv.org/abs/1503.08043} {arXiv:1503.08043 [hep-th]} \BibitemShut
  {NoStop}%
\bibitem [{\citenamefont {Lee}\ \emph {et~al.}(2016)\citenamefont {Lee},
  \citenamefont {Baumann},\ and\ \citenamefont {Pimentel}}]{Lee:2016vti}%
  \BibitemOpen
  \bibfield  {author} {\bibinfo {author} {\bibfnamefont {H.}~\bibnamefont
  {Lee}}, \bibinfo {author} {\bibfnamefont {D.}~\bibnamefont {Baumann}}, \ and\
  \bibinfo {author} {\bibfnamefont {G.~L.}\ \bibnamefont {Pimentel}},\ }\href
  {\doibase 10.1007/JHEP12(2016)040} {\bibfield  {journal} {\bibinfo  {journal}
  {JHEP}\ }\textbf {\bibinfo {volume} {12}},\ \bibinfo {pages} {040} (\bibinfo
  {year} {2016})},\ \Eprint {http://arxiv.org/abs/1607.03735} {arXiv:1607.03735
  [hep-th]} \BibitemShut {NoStop}%
\bibitem [{\citenamefont {Chen}\ \emph
  {et~al.}(2017{\natexlab{a}})\citenamefont {Chen}, \citenamefont {Wang},\ and\
  \citenamefont {Xianyu}}]{Chen:2016uwp}%
  \BibitemOpen
  \bibfield  {author} {\bibinfo {author} {\bibfnamefont {X.}~\bibnamefont
  {Chen}}, \bibinfo {author} {\bibfnamefont {Y.}~\bibnamefont {Wang}}, \ and\
  \bibinfo {author} {\bibfnamefont {Z.-Z.}\ \bibnamefont {Xianyu}},\ }\href
  {\doibase 10.1103/PhysRevLett.118.261302} {\bibfield  {journal} {\bibinfo
  {journal} {Phys. Rev. Lett.}\ }\textbf {\bibinfo {volume} {118}},\ \bibinfo
  {pages} {261302} (\bibinfo {year} {2017}{\natexlab{a}})},\ \Eprint
  {http://arxiv.org/abs/1610.06597} {arXiv:1610.06597 [hep-th]} \BibitemShut
  {NoStop}%
\bibitem [{\citenamefont {Chen}\ \emph
  {et~al.}(2017{\natexlab{b}})\citenamefont {Chen}, \citenamefont {Wang},\ and\
  \citenamefont {Xianyu}}]{Chen:2016hrz}%
  \BibitemOpen
  \bibfield  {author} {\bibinfo {author} {\bibfnamefont {X.}~\bibnamefont
  {Chen}}, \bibinfo {author} {\bibfnamefont {Y.}~\bibnamefont {Wang}}, \ and\
  \bibinfo {author} {\bibfnamefont {Z.-Z.}\ \bibnamefont {Xianyu}},\ }\href
  {\doibase 10.1007/JHEP04(2017)058} {\bibfield  {journal} {\bibinfo  {journal}
  {JHEP}\ }\textbf {\bibinfo {volume} {04}},\ \bibinfo {pages} {058} (\bibinfo
  {year} {2017}{\natexlab{b}})},\ \Eprint {http://arxiv.org/abs/1612.08122}
  {arXiv:1612.08122 [hep-th]} \BibitemShut {NoStop}%
\bibitem [{\citenamefont {Meerburg}\ \emph {et~al.}(2019)\citenamefont
  {Meerburg} \emph {et~al.}}]{Meerburg:2019qqi}%
  \BibitemOpen
  \bibfield  {author} {\bibinfo {author} {\bibfnamefont {P.~D.}\ \bibnamefont
  {Meerburg}} \emph {et~al.},\ }\href@noop {} {\  (\bibinfo {year} {2019})},\
  \Eprint {http://arxiv.org/abs/1903.04409} {arXiv:1903.04409 [astro-ph.CO]}
  \BibitemShut {NoStop}%
\bibitem [{\citenamefont {Meerburg}\ \emph {et~al.}(2017)\citenamefont
  {Meerburg}, \citenamefont {M{\"u}nchmeyer}, \citenamefont {Mu{\~n}oz},\ and\
  \citenamefont {Chen}}]{Meerburg:2016zdz}%
  \BibitemOpen
  \bibfield  {author} {\bibinfo {author} {\bibfnamefont {P.~D.}\ \bibnamefont
  {Meerburg}}, \bibinfo {author} {\bibfnamefont {M.}~\bibnamefont
  {M{\"u}nchmeyer}}, \bibinfo {author} {\bibfnamefont {J.~B.}\ \bibnamefont
  {Mu{\~n}oz}}, \ and\ \bibinfo {author} {\bibfnamefont {X.}~\bibnamefont
  {Chen}},\ }\href {\doibase 10.1088/1475-7516/2017/03/050} {\bibfield
  {journal} {\bibinfo  {journal} {JCAP}\ }\textbf {\bibinfo {volume} {1703}},\
  \bibinfo {pages} {050} (\bibinfo {year} {2017})},\ \Eprint
  {http://arxiv.org/abs/1610.06559} {arXiv:1610.06559 [astro-ph.CO]}
  \BibitemShut {NoStop}%
\bibitem [{\citenamefont {Moradinezhad~Dizgah}\ \emph
  {et~al.}(2018)\citenamefont {Moradinezhad~Dizgah}, \citenamefont {Lee},
  \citenamefont {Mu{\~n}oz},\ and\ \citenamefont
  {Dvorkin}}]{MoradinezhadDizgah:2018ssw}%
  \BibitemOpen
  \bibfield  {author} {\bibinfo {author} {\bibfnamefont {A.}~\bibnamefont
  {Moradinezhad~Dizgah}}, \bibinfo {author} {\bibfnamefont {H.}~\bibnamefont
  {Lee}}, \bibinfo {author} {\bibfnamefont {J.~B.}\ \bibnamefont {Mu{\~n}oz}},
  \ and\ \bibinfo {author} {\bibfnamefont {C.}~\bibnamefont {Dvorkin}},\ }\href
  {\doibase 10.1088/1475-7516/2018/05/013} {\bibfield  {journal} {\bibinfo
  {journal} {JCAP}\ }\textbf {\bibinfo {volume} {1805}},\ \bibinfo {pages}
  {013} (\bibinfo {year} {2018})},\ \Eprint {http://arxiv.org/abs/1801.07265}
  {arXiv:1801.07265 [astro-ph.CO]} \BibitemShut {NoStop}%
\bibitem [{\citenamefont {Kogai}\ \emph {et~al.}(2021)\citenamefont {Kogai},
  \citenamefont {Akitsu}, \citenamefont {Schmidt},\ and\ \citenamefont
  {Urakawa}}]{Kogai:2020vzz}%
  \BibitemOpen
  \bibfield  {author} {\bibinfo {author} {\bibfnamefont {K.}~\bibnamefont
  {Kogai}}, \bibinfo {author} {\bibfnamefont {K.}~\bibnamefont {Akitsu}},
  \bibinfo {author} {\bibfnamefont {F.}~\bibnamefont {Schmidt}}, \ and\
  \bibinfo {author} {\bibfnamefont {Y.}~\bibnamefont {Urakawa}},\ }\href
  {\doibase 10.1088/1475-7516/2021/03/060} {\bibfield  {journal} {\bibinfo
  {journal} {JCAP}\ }\textbf {\bibinfo {volume} {03}},\ \bibinfo {pages} {060}
  (\bibinfo {year} {2021})},\ \Eprint {http://arxiv.org/abs/2009.05517}
  {arXiv:2009.05517 [astro-ph.CO]} \BibitemShut {NoStop}%
\bibitem [{\citenamefont {Wang}\ \emph {et~al.}(2022)\citenamefont {Wang},
  \citenamefont {Xianyu},\ and\ \citenamefont {Zhong}}]{Wang:2021qez}%
  \BibitemOpen
  \bibfield  {author} {\bibinfo {author} {\bibfnamefont {L.-T.}\ \bibnamefont
  {Wang}}, \bibinfo {author} {\bibfnamefont {Z.-Z.}\ \bibnamefont {Xianyu}}, \
  and\ \bibinfo {author} {\bibfnamefont {Y.-M.}\ \bibnamefont {Zhong}},\ }\href
  {\doibase 10.1007/JHEP02(2022)085} {\bibfield  {journal} {\bibinfo  {journal}
  {JHEP}\ }\textbf {\bibinfo {volume} {02}},\ \bibinfo {pages} {085} (\bibinfo
  {year} {2022})},\ \Eprint {http://arxiv.org/abs/2109.14635} {arXiv:2109.14635
  [hep-ph]} \BibitemShut {NoStop}%
\bibitem [{\citenamefont {Tong}\ \emph {et~al.}(2021)\citenamefont {Tong},
  \citenamefont {Wang},\ and\ \citenamefont {Zhu}}]{Tong:2021wai}%
  \BibitemOpen
  \bibfield  {author} {\bibinfo {author} {\bibfnamefont {X.}~\bibnamefont
  {Tong}}, \bibinfo {author} {\bibfnamefont {Y.}~\bibnamefont {Wang}}, \ and\
  \bibinfo {author} {\bibfnamefont {Y.}~\bibnamefont {Zhu}},\ }\href@noop {} {\
   (\bibinfo {year} {2021})},\ \Eprint {http://arxiv.org/abs/2112.03448}
  {arXiv:2112.03448 [hep-th]} \BibitemShut {NoStop}%
\bibitem [{\citenamefont {Ade}\ \emph {et~al.}(2021)\citenamefont {Ade} \emph
  {et~al.}}]{BICEP:2021xfz}%
  \BibitemOpen
  \bibfield  {author} {\bibinfo {author} {\bibfnamefont {P.~A.~R.}\
  \bibnamefont {Ade}} \emph {et~al.} (\bibinfo {collaboration} {BICEP, Keck}),\
  }\href {\doibase 10.1103/PhysRevLett.127.151301} {\bibfield  {journal}
  {\bibinfo  {journal} {Phys. Rev. Lett.}\ }\textbf {\bibinfo {volume} {127}},\
  \bibinfo {pages} {151301} (\bibinfo {year} {2021})},\ \Eprint
  {http://arxiv.org/abs/2110.00483} {arXiv:2110.00483 [astro-ph.CO]}
  \BibitemShut {NoStop}%
\bibitem [{\citenamefont {An}\ \emph {et~al.}(2018)\citenamefont {An},
  \citenamefont {McAneny}, \citenamefont {Ridgway},\ and\ \citenamefont
  {Wise}}]{An:2017hlx}%
  \BibitemOpen
  \bibfield  {author} {\bibinfo {author} {\bibfnamefont {H.}~\bibnamefont
  {An}}, \bibinfo {author} {\bibfnamefont {M.}~\bibnamefont {McAneny}},
  \bibinfo {author} {\bibfnamefont {A.~K.}\ \bibnamefont {Ridgway}}, \ and\
  \bibinfo {author} {\bibfnamefont {M.~B.}\ \bibnamefont {Wise}},\ }\href
  {\doibase 10.1007/JHEP06(2018)105} {\bibfield  {journal} {\bibinfo  {journal}
  {JHEP}\ }\textbf {\bibinfo {volume} {06}},\ \bibinfo {pages} {105} (\bibinfo
  {year} {2018})},\ \Eprint {http://arxiv.org/abs/1706.09971} {arXiv:1706.09971
  [hep-ph]} \BibitemShut {NoStop}%
\bibitem [{\citenamefont {Iyer}\ \emph {et~al.}(2018)\citenamefont {Iyer},
  \citenamefont {Pi}, \citenamefont {Wang}, \citenamefont {Wang},\ and\
  \citenamefont {Zhou}}]{Iyer:2017qzw}%
  \BibitemOpen
  \bibfield  {author} {\bibinfo {author} {\bibfnamefont {A.~V.}\ \bibnamefont
  {Iyer}}, \bibinfo {author} {\bibfnamefont {S.}~\bibnamefont {Pi}}, \bibinfo
  {author} {\bibfnamefont {Y.}~\bibnamefont {Wang}}, \bibinfo {author}
  {\bibfnamefont {Z.}~\bibnamefont {Wang}}, \ and\ \bibinfo {author}
  {\bibfnamefont {S.}~\bibnamefont {Zhou}},\ }\href {\doibase
  10.1088/1475-7516/2018/01/041} {\bibfield  {journal} {\bibinfo  {journal}
  {JCAP}\ }\textbf {\bibinfo {volume} {1801}},\ \bibinfo {pages} {041}
  (\bibinfo {year} {2018})},\ \Eprint {http://arxiv.org/abs/1710.03054}
  {arXiv:1710.03054 [hep-th]} \BibitemShut {NoStop}%
\bibitem [{\citenamefont {Chen}\ \emph
  {et~al.}(2017{\natexlab{c}})\citenamefont {Chen}, \citenamefont {Wang},\ and\
  \citenamefont {Xianyu}}]{Chen:2017ryl}%
  \BibitemOpen
  \bibfield  {author} {\bibinfo {author} {\bibfnamefont {X.}~\bibnamefont
  {Chen}}, \bibinfo {author} {\bibfnamefont {Y.}~\bibnamefont {Wang}}, \ and\
  \bibinfo {author} {\bibfnamefont {Z.-Z.}\ \bibnamefont {Xianyu}},\ }\href
  {\doibase 10.1088/1475-7516/2017/12/006} {\bibfield  {journal} {\bibinfo
  {journal} {JCAP}\ }\textbf {\bibinfo {volume} {1712}},\ \bibinfo {pages}
  {006} (\bibinfo {year} {2017}{\natexlab{c}})},\ \Eprint
  {http://arxiv.org/abs/1703.10166} {arXiv:1703.10166 [hep-th]} \BibitemShut
  {NoStop}%
\bibitem [{\citenamefont {Wang}\ and\ \citenamefont
  {Xianyu}(2020)}]{Wang:2019gbi}%
  \BibitemOpen
  \bibfield  {author} {\bibinfo {author} {\bibfnamefont {L.-T.}\ \bibnamefont
  {Wang}}\ and\ \bibinfo {author} {\bibfnamefont {Z.-Z.}\ \bibnamefont
  {Xianyu}},\ }\href {\doibase 10.1007/JHEP02(2020)044} {\bibfield  {journal}
  {\bibinfo  {journal} {JHEP}\ }\textbf {\bibinfo {volume} {02}},\ \bibinfo
  {pages} {044} (\bibinfo {year} {2020})},\ \Eprint
  {http://arxiv.org/abs/1910.12876} {arXiv:1910.12876 [hep-ph]} \BibitemShut
  {NoStop}%
\bibitem [{\citenamefont {Chen}\ \emph {et~al.}(2016)\citenamefont {Chen},
  \citenamefont {Namjoo},\ and\ \citenamefont {Wang}}]{Chen:2015lza}%
  \BibitemOpen
  \bibfield  {author} {\bibinfo {author} {\bibfnamefont {X.}~\bibnamefont
  {Chen}}, \bibinfo {author} {\bibfnamefont {M.~H.}\ \bibnamefont {Namjoo}}, \
  and\ \bibinfo {author} {\bibfnamefont {Y.}~\bibnamefont {Wang}},\ }\href
  {\doibase 10.1088/1475-7516/2016/02/013} {\bibfield  {journal} {\bibinfo
  {journal} {JCAP}\ }\textbf {\bibinfo {volume} {1602}},\ \bibinfo {pages}
  {013} (\bibinfo {year} {2016})},\ \Eprint {http://arxiv.org/abs/1509.03930}
  {arXiv:1509.03930 [astro-ph.CO]} \BibitemShut {NoStop}%
\bibitem [{\citenamefont {Lu}\ \emph {et~al.}(2021)\citenamefont {Lu},
  \citenamefont {Reece},\ and\ \citenamefont {Xianyu}}]{Lu:2021wxu}%
  \BibitemOpen
  \bibfield  {author} {\bibinfo {author} {\bibfnamefont {Q.}~\bibnamefont
  {Lu}}, \bibinfo {author} {\bibfnamefont {M.}~\bibnamefont {Reece}}, \ and\
  \bibinfo {author} {\bibfnamefont {Z.-Z.}\ \bibnamefont {Xianyu}},\ }\href
  {\doibase 10.1007/JHEP12(2021)098} {\bibfield  {journal} {\bibinfo  {journal}
  {JHEP}\ }\textbf {\bibinfo {volume} {12}},\ \bibinfo {pages} {098} (\bibinfo
  {year} {2021})},\ \Eprint {http://arxiv.org/abs/2108.11385} {arXiv:2108.11385
  [hep-ph]} \BibitemShut {NoStop}%
\bibitem [{\citenamefont {Chen}\ \emph {et~al.}(2019)\citenamefont {Chen},
  \citenamefont {Loeb},\ and\ \citenamefont {Xianyu}}]{Chen:2018cgg}%
  \BibitemOpen
  \bibfield  {author} {\bibinfo {author} {\bibfnamefont {X.}~\bibnamefont
  {Chen}}, \bibinfo {author} {\bibfnamefont {A.}~\bibnamefont {Loeb}}, \ and\
  \bibinfo {author} {\bibfnamefont {Z.-Z.}\ \bibnamefont {Xianyu}},\ }\href
  {\doibase 10.1103/PhysRevLett.122.121301} {\bibfield  {journal} {\bibinfo
  {journal} {Phys. Rev. Lett.}\ }\textbf {\bibinfo {volume} {122}},\ \bibinfo
  {pages} {121301} (\bibinfo {year} {2019})},\ \Eprint
  {http://arxiv.org/abs/1809.02603} {arXiv:1809.02603 [astro-ph.CO]}
  \BibitemShut {NoStop}%
\bibitem [{\citenamefont {Wang}\ \emph {et~al.}(2020)\citenamefont {Wang},
  \citenamefont {Wang},\ and\ \citenamefont {Zhu}}]{Wang:2020aqc}%
  \BibitemOpen
  \bibfield  {author} {\bibinfo {author} {\bibfnamefont {Y.}~\bibnamefont
  {Wang}}, \bibinfo {author} {\bibfnamefont {Z.}~\bibnamefont {Wang}}, \ and\
  \bibinfo {author} {\bibfnamefont {Y.}~\bibnamefont {Zhu}},\ }\href {\doibase
  10.1088/1475-7516/2020/11/026} {\bibfield  {journal} {\bibinfo  {journal}
  {JCAP}\ }\textbf {\bibinfo {volume} {11}},\ \bibinfo {pages} {026} (\bibinfo
  {year} {2020})},\ \Eprint {http://arxiv.org/abs/2007.09677} {arXiv:2007.09677
  [hep-th]} \BibitemShut {NoStop}%
\end{thebibliography}%
 
\begin{appendix}

\section*{Appendix}

\noindent\textbf{The scalar tilt.} In the main text we have showed the correction to the scalar tilt $\Delta n_s$ due to the time-dependent mass, in Fig.\ \ref{fig_ns}. To get this figure, we have numerically evolved the mode equations (\ref{eq_EoM}). The Bunch-Davies initial condition is imposed as usual, at an initial time when all relevant modes are well inside the horizon. Then, we collect the values of the resulting mode function at a final time where all relevant modes are well outside the horizon. With these mode functions, we get a numerical result for the power spectrum $P_\zeta$ in (\ref{eq_PS}) for a range of comoving momentum $k$. We then fit the numerical result of $\log P_\zeta$ by a linear function $X\log k + Y$ in $\log k$. The scalar tilt $n_s$ can thus be extracted from the slope of this fitted linear function, $n_s-1=X$. We have also checked numerically that the running of the scalar tilt is around or below $\order{10^{-2}}$. Therefore, the linear function $X\log k + Y$ is a good fit of the numerical power spectrum for the parameter space of interest. 

We emphasize again that Fig.\ \ref{fig_ns} shows the corrections to the $n_s$ from the time-dependent mass $m(\tau)$ alone. The complete result of $n_s$ should also include the slow-roll corrections. The slow-roll correction can be included by rewriting the mode equations (\ref{eq_EoM}) with a general slow-roll background $a(\tau)$ instead of the slow-roll limit $a(\tau)=-1/(H\tau)$. Numerically evolving this slow-roll corrected mode equations can give the full result of $n_s$. In this Letter, we chose not to include the slow-roll correction in the mode equation, in order to isolate and highlight the effect of the time-dependent mass. 

To clarify the relation between our result and the scalar tilt from a constant-mass spectator computed in \cite{An:2017hlx}, we write the power spectrum $P_\zeta$ in the following way:
\begin{align}
  P_\zeta = P_\zeta(m/H,\mu/H),
\end{align}
where $m$ is the spectator's mass and $\mu$ is its two-point coupling with the inflaton. The scale ($k$) dependence in $P_\zeta$ is obtained from the time ($\tau$) dependence of $m/H$ and $\mu/H$, by evaluating $m/H$ and $\mu/H$ at the time of horizon exit, $\tau\simeq-1/k$. In \cite{An:2017hlx}, the scalar tilt is obtained by including the time dependence in $H$:
\begin{align}
  P_\zeta = P_\zeta\Big(\FR{m}{H(\tau)},\FR{\mu}{H(\tau)}\Big),
\end{align}
while in this Letter, we consider the time dependence in $m$, and switched off the time dependence in $H$:
\begin{align}
  P_\zeta = P_\zeta\Big(\FR{m(\tau)}{H},\FR{\mu}{H}\Big).
\end{align}

\noindent\textbf{Numerical implementation.} In the main text we presented the numerical result of the in-in integral (\ref{eq_bispect}) which computes the Feynman diagram in (\ref{eq_3ptdiagram}). Here we spell out the procedure to carry out this numerical calculation, essentially following the treatment introduced in \cite{An:2017hlx}. 

The computation begins with the usual canonical quantization of the Fourier modes $\varphi_\mb{k}(\tau)$ and $\si_\mb{k}(\tau)$, which can be expanded in terms of a set of creation and annihilation operators $a_\mb{k}$, $a^\dag_\mb{k}$, $b_\mb{k}$, $b^\dag_\mb{k}$, as follows.
\begin{align}
  &\varphi_\mb{k}(\tau)=\big[\varphi_k^{(1)}a_\mb{k}(\tau) +\varphi_k^{(2)}b_\mb{k}(\tau)\big]e^{\ii\mb k\cdot\mb x}+\text{h.c.},\\
  &\si_\mb{k}(\tau)=\big[\si_k^{(1)}a_\mb{k}(\tau) +\si_k^{(2)}b_\mb{k}(\tau)\big]e^{\ii\mb k\cdot\mb x}+\text{h.c.},
\end{align}
where the variables $\varphi_k^{(n)}$ and $\si_k^{(n)}$~($n=1,2$) with italic subscript $k$ are mode functions, satisfying the same set of equations (\ref{eq_EoM}) as the field operators. Here we have included $a_\mb{k}$ and $b_\mb{k}$ in both field operators, allowing for the mixing between the two fields. To solve the mode equations (\ref{eq_EoM}), we impose the usual Bunch-Davies initial condition in the early-time limit $\tau\to-\infty$ with proper normalizations:
\begin{align}
  &\varphi_{k}^{(n)}=\FR{H}{2k^{3/2}}(-k\tau)^{1-(-1)^n \ii\mu/(2H)}e^{-\ii k\tau},\\
  &\si_k^{(n)}=(-1)^{n-1}\ii\varphi_k^{(n)}.
\end{align}
The unusual power dependence on $\tau$ is due to the mixing of the two modes. Then, in principle, we can numerically solve (\ref{eq_EoM}) with initial conditions spelled out above. However, one can anticipate that the integrand of (\ref{eq_bispect}) becomes highly oscillatory in the early-time limit $\tau\to-\infty$, making the numerical computation difficult. Therefore, we Wick-rotate the time variable $\tau\to\pm\ii\tau$ such that the mode function is always exponentially damped when $\tau\to-\infty$. Note that this specification of the contour rotation is consistent with the standard $\ii\ep$-prescription, which is ultimately from selecting the correct vacuum initial state. 

\begin{figure}[t]
  \centering
  \includegraphics[width=0.4\textwidth]{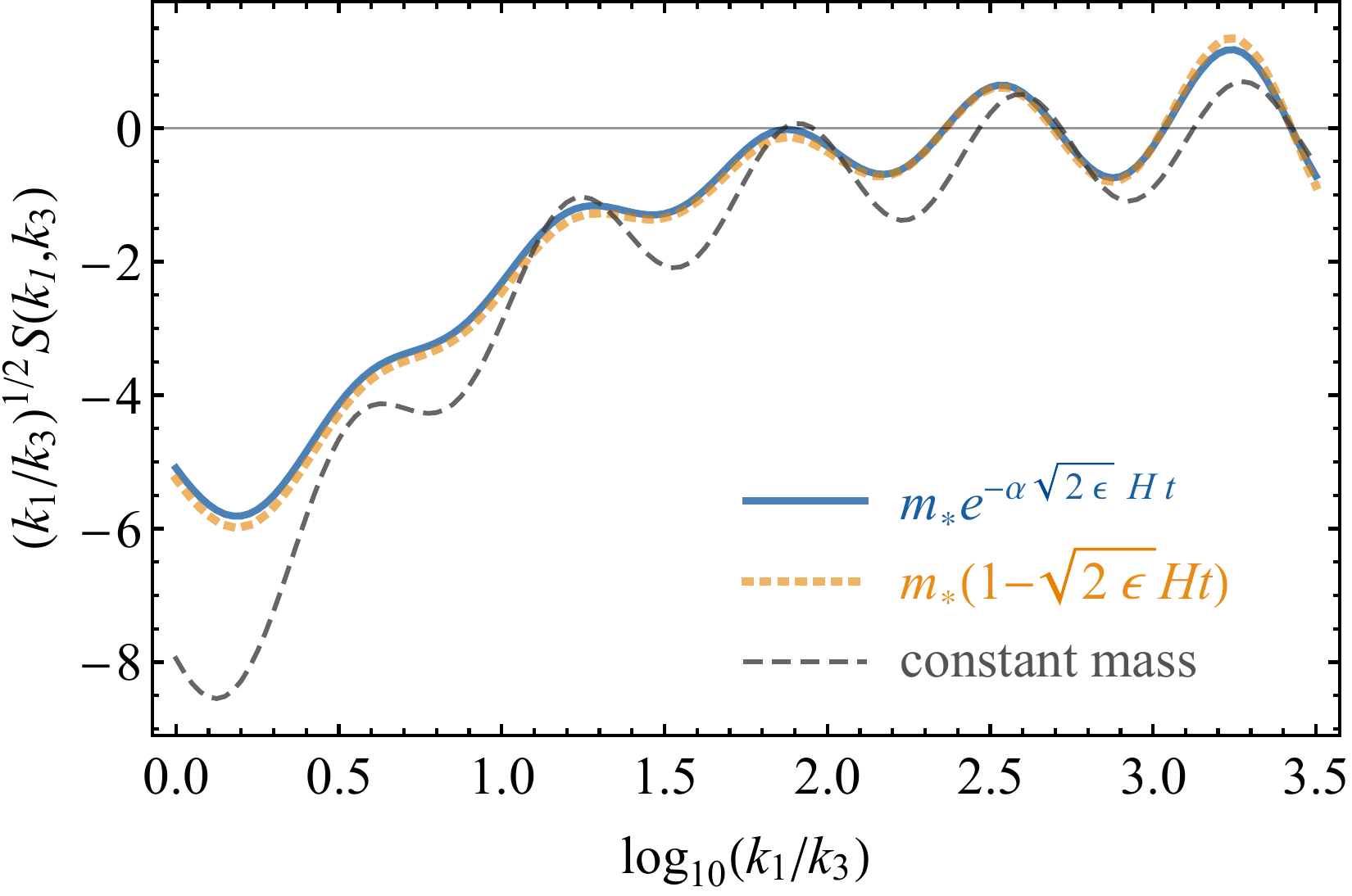}
  \caption{The shape function (\ref{eq_shape}) with two parameterizations of the mass $m_\text{eff}(t)$, plotted with the constant-mass shape function. The blue curve corresponds to (\ref{msigma}) and the orange curve corresponds to (\ref{eq_massalter}). For this plot we choose $\al=1$ and $16\ep=0.005$, and other parameters are chosen to be the same with in Fig.\;\ref{fig_signal}. }
  \label{fig_altersignal}
\end{figure}

There is one more complication in the numerical computation: After the Wick rotation, the two oscillatory branches of the mode function become exponential functions of the imaginary time $\ii\tau$, either growing or dying exponentially fast as $\tau\to 0$. In the numerical calculation, we need to specify the initial condition at some large $-\tau_i$, and then evolve the equation numerically towards $\tau=0$. The existence of an exponentially growing branch means that we have to specify the initial condition exponentially precisely in order to get a result of any reasonable precision. This is virtually impossible. However, there is a simple way out: Since we know that the mode functions would be essentially exponential functions in the early-time limit, we can simply factor this exponential  out before the numerical computation. This motivates us to rewrite 
\begin{align}
  &\varphi_k(\tau)=A_k(\tau) e^{-\ii k\tau},
  &&\si_k(\tau)=B_k(\tau)e^{-\ii k\tau}.
\end{align}
Then the functions $A_k$ and $B_k$ would have no dangerous exponential dependence on $\tau$, making the numerical computation safe. It is simple to get a pair of equations for $A_k$ and $B_k$, written with a dimensionless variable $z=-\ii k\tau$:
\begin{align}
   A_k''+ \FR{2(z-1)}{z} A_k'-\FR{2}{z}A_k-\FR{\mu}{z}B_k'- \FR{\mu(z-3)}{z^2} B_k=0,\n\\
   B_k''+ \FR{2(z-1)}{z} B_k'+ \frac{m^2(z)-2z}{z^2} B_k+\FR{\mu}{z}(A_k'+ A_k)=0,\n
\end{align}
where $m^2(z)=m_*^2(z^2/k^2)^{\al\sqrt{2\ep}}e^{\ii\pi\al\sqrt{2\ep}}$ is the time-dependent mass after the Wick rotation, and the prime denotes the derivative with respect to $z$. 

With the above numerical procedure, the in-in integral (\ref{eq_bispect}) can be carried out directly in terms of the mode functions:
\begin{align}
&\la \varphi_{\mb k_1}\varphi_{\mb k_2}\varphi_{\mb k_3}\ra'=\; 2\lam_3\,\text{Im}\int_{-\infty}^0 \di\tau\, a^4\n\\
& \times\prod_{i=1}^3\Big[\si_{k_i}^{(1)}(\tau)\varphi_{k_i}^{(1)}(0)+\si_{k_i}^{(2)}(\tau)\varphi_{k_i}^{(2)}(0)\Big],
\end{align}
where we have taken $\tau_f=0$.

\noindent\textbf{Alternative mass ansatz.}
In the main text we focus on an exponential ansatz for the time-dependent mass of the spectator field. Here, instead of (\ref{msigma}), we consider an alternative parametrization of the time dependence in the mass:
\bge
\label{eq_massalter}
  m_\text{eff}(t)=m_*(1-\sqrt{2\ep_*}H_* t),
\ede
which could come from a direct coupling $\lam\phi^2\si^2$. Due to the smallness of $\ep_*$, we can expect that the signal from this direct coupling would be similar to the exponential ansatz (\ref{msigma}) for at least a range of $k$. In Fig.\;\ref{fig_altersignal} we show both parameterizations of the mass and it is clear that the two shape functions with time-dependent mass are very close to each other, and both show clear deviations from the constant-mass signal. Therefore, the main conclusion of our paper is insensitive to the particular type of direct coupling between the spectator field and the inflaton.

\end{appendix}

\end{document}